\DocumentMetadata{}

\documentclass[sigconf]{acmart}

\usepackage{xcolor}
\usepackage{pdflscape}
\usepackage{multirow}

\usepackage{mdframed}
\AtBeginDocument{%
  \providecommand\BibTeX{{%
    \normalfont B\kern-0.5em{\scshape i\kern-0.25em b}\kern-0.8em\TeX}}}

\copyrightyear{2024} 
\acmYear{2024} 
\setcopyright{acmlicensed}
\acmConference[SBQS25]{Brazilian Symposium on Software Quality}{Nov 04--07, 2025}{São José dos Campos, SP}

\setcopyright{none}
\renewcommand\footnotetextcopyrightpermission[1]{}

\setcopyright{none}
\renewcommand\footnotetextcopyrightpermission[1]{}

\acmDOI{XXXXXXX.XXXXXXX}

\makeatletter
\def\@secfont{\bfseries\Large\section@raggedright}
\makeatother

\begin{document}
\renewcommand\abstractname{ABSTRACT}
\renewcommand\keywordsname{KEYWORDS}
\renewcommand\refname{REFERENCES}

\title[ ]{DIP-AI: A Discovery Framework for AI Innovation Projects}
\renewcommand{\shorttitle}{DIP-AI: A Discovery Framework for AI Innovation Projects}



\author{Mariana Crisostomo Martins}
\affiliation{%
  \institution{Instituto de Informática Federal\\Universidade Federal de Goiás}
  \city{Goiânia, GO}
  \country{Brazil}}
\email{maricrisostomo.martins@gmail.com}
\author{Lucas Elias Cardoso Rocha}
\affiliation{%
  \institution{Centro de Excelência em Inteligência Artificial}
  \city{Goiânia, GO}
  \country{Brazil}}
\email{lucaseliascrocha@gmail.com}
\author{Lucas Cordeiro Romão}
\affiliation{%
  \institution{Departamento de Informática\\Pontifícia Católica Universidade do Rio de Janeiro}
  \city{Rio de Janeiro, RJ}
  \country{Brazil}}
\email{lc.romao98@gmail.com}
\author{Taciana Novo Kudo}
\affiliation{%
  \institution{Instituto de Informática Federal\\Universidade Federal de Goiás}
  \city{Goiânia, GO}
  \country{Brazil}}
\email{taciana@ufg.br}
\author{Marcos Kalinowski}
\affiliation{%
  \institution{Departamento de Informática\\Pontifícia Católica Universidade do Rio de Janeiro}
  \city{Rio de Janeiro, RJ}
  \country{Brazil}}
\email{kalinowski@inf.puc-rio.br}
\author{Renato de Freitas Bulcão-Neto}
\affiliation{%
  \institution{Instituto de Informática Federal\\Universidade Federal de Goiás}
    \city{Goiânia, GO}
  \country{Brazil}}
\email{rbulcao@ufg.br}



\renewcommand{\shortauthors}{Martins et al.}

\begin{abstract}

Despite the increasing development of Artificial Intelligence (AI) systems, Requirements Engineering (RE) activities face challenges in this new data-intensive paradigm. We identified a lack of support for problem discovery within AI innovation projects. To address this, we propose and evaluate DIP-AI, a discovery framework tailored to guide early-stage exploration in such initiatives. Based on a literature review, our solution proposal combines elements of ISO 12207, 5338, and Design Thinking to support the discovery of AI innovation projects, aiming at promoting higher quality deliveries and stakeholder satisfaction. We evaluated DIP-AI in an industry-academia collaboration (IAC) case study of an AI innovation project, in which participants applied DIP-AI to the discovery phase in practice and provided their perceptions about the approach's problem discovery capability, acceptance, and suggestions. The results indicate that DIP-AI is relevant and useful, particularly in facilitating problem discovery in AI projects. This research contributes to academia by sharing DIP-AI as a framework for AI problem discovery. For industry, we discuss the use of this framework in a real IAC program that develops AI innovation projects. 
\end{abstract}


       


\keywords{Requirements Engineering, Discovery, AI Systems, Innovation Projects.}



\maketitle

\section{Introduction}

The increasing adoption of Artificial Intelligence (AI) components in software raises concerns regarding the quality of these components. Many factors 
have contributed to the widespread integration of Machine Learning (ML) components into traditional software. Examples of such systems can be found in healthcare \cite{jiang2017artificial}, finance \cite{goodell2021artificial}, education \cite{kuvcak2018machine}, and others.

Difficulty in communication and high customer expectations are aspects commonly addressed in Requirements Engineering (RE) and identified as particularly challenging in ML projects \cite{IshikawaYoshioka}. These systems often involve ambiguous requirements, emergent quality attributes such as explainability, and inherently iterative and experimental development cycles. Despite its critical role in quality assurance, current RE practices offer limited guidance for ML-based systems \cite{ahmad2022mapeamento} 

Martins et al. \cite{Martinscibse} conducted a study on the state of RE practice in ML projects within an innovation-driven environment, focusing on research, development, and innovation (RDI) projects. Through interviews with project coordinators, they identified several challenges, including problem comprehension, deriving insights from data, maintaining effective communication, and aligning technical and domain knowledge to manage feasibility constraints challenges. These findings are aligned with the findings from the survey by Alves \textit{et al.} \cite{alvesAPKalinowski-statusquo}, which also identified difficulties in problem understanding, managing customer expectations, and aligning requirements with data as key challenges for the improvement of the quality of these systems. 

The development with AI components represents a paradigm change compared to traditional software development. AI systems present AI models that have data-based behavior while in traditional software the behavior is specified according to user needs and business rules. RE is a challenge in this scenario, being critical and directly affecting the quality of these systems \cite{giray2021software, villamizar2021requirements}. The traditional practices of RE are not well aligned with AI development \cite{giray2021software, zaharia2018accelerating,hu2020towards}. 

Marban \textit{et al.} \cite{Marban09} present a comparison between CRISP-DM \cite{CRISP-DM} and the methodologies proposed by Fayyad \textit{et al.} \cite{fayyad}, Cabena \textit{et al.} \cite{cabena}, and Anand and Bucher \cite{anand98}. These methodologies have in common a \textit{Discovery} phase, comprising the identification of human resources, understanding the business, determining business objectives, and developing and understanding the application domain. Although they are aligned with the objectives of ISO 12207, there is no clear mapping to the aspects of the ISO requirements. The identification and characterization of the business becomes a challenge to be addressed by the project team, usually by the data scientist. In practice, this stakeholder typically conducts RE-related activities \cite{nascimento2020, alvesAPKalinowski-statusquo}. However, these activities would be better developed by a specialist with the support of data scientists and other stakeholders ~\cite{villamizar2024identifying}. In fact, to identify a business and characterize it in search of points that reinforce the need for AI solutions for the business, it is necessary to have a direction to support this activity, together with other stakeholders.

This paper proposes a framework for discovery in AI innovation projects that helps to understand the business, its objectives, and its problems, as well as to prepare for the project definition. This framework guides requirements analysts in understanding the organization, problems, data, and identifying characteristics and their distribution, which is essential to achieve a high-quality AI model \cite{Wan2020}. This solution facilitates understanding of the context and domain \cite{heyn2021requirement}, enhances stakeholder collaboration \cite{IshikawaYoshioka}, and enables anticipation and structured discussions on model degradation \cite{vogelsang2019requirements}, perception regarding quality, non-functional requirements (NFR), and trade-offs \cite{ahmad2022mapeamento,villamizar2022perspectivebased,habibullah2023}. We developed the proposed framework based on the Technology Transfer Model (TTM)~\cite{pfleeger}, incorporating foundational elements from ISO 12207, 5338, and Design Thinking.

To assess the effectiveness of the proposed framework and gather insights for its improvement, we conducted a case study following the guidelines outlined by Runeson and Höst \cite{Runeson2008}. The study was carried out within the context of a research and development (R\&D) initiative, independent of the authors' institution, and involved a real industrial partner engaged in the development of a minimum viable product (MVP). The evaluation of the DIP-AI framework was structured as an evaluative case study, focusing on the perspectives of participants involved in the problem discovery phase.

We designed a structured questionnaire aimed to capture participants’ perceptions regarding the problem discovery capability, usability, and utility of the framework, with the quantitative items grounded in the constructs of the Technology Acceptance Model (TAM) \cite{tam3}. We also collected suggestions with additional open-text answer options.

As part of the case study, DIP-AI was successfully applied in practice, helping to understand the business objectives and problems, providing the basis for subsequently defining an MVP of an AI product that was successfully developed and deployed at the industrial partner. The main findings of the quantitative analysis indicate that the proposed approach is useful, relevant, and effective in supporting the discovery of AI innovation projects. Qualitative analysis of participants' feedback revealed additional suggestions for enhancement, particularly regarding the need for additional support for applying the framework.

This paper is organized as follows. Section 2 describes the related work. Section 3 presents the background of the proposed framework. Section 4 presents the proposed framework and describes the method for defining the framework. Section 5 describes the study case design. Section 6 provides the study case results. Section 7 presents the discussion of the results. Section 8 discusses threats to validity. Finally, Section 9 concludes this work, highlighting its contributions and future work.

\section{Related Work}
\label{sec:relatedwork}

This section reviews related approaches to RE for AI systems. While existing works define requirements and perspectives (e.g., PerSpecML \cite{villamizar2024identifying}) or combine methodologies like CRISP-DM, DT, and Lean \cite{ahmed}, none focus specifically on a structured Discovery phase for AI projects.

PerSpecML organizes RE-related concerns into five perspectives (goals, user experience, infrastructure, model, and data) and supports their specification through guided questions. Our approach complements it by preceding the specification with structured problem discovery and feasibility analysis. Ahmed \textit{et al.} propose an integrated methodology (business, data, product), but do not address specific aspects of development defined in the RE process and relevant for quality development.

The AI Canvas \cite{dewalt}, the ML Canvas \cite{dorard2015}, and Deep Learning (DL) Canvas \cite{perez} are canvases that present fields with context application, key evaluation metrics, features, model development, and data aspects (distribution, source, and logistics). All of these fields are very general and, therefore, depend on the expertise of the filler to obtain the necessary information. In addition, aspects regarding the decision of which problem(s) will be solved, data availability, model consumption method, etc. are not addressed in the canvas and can strongly impact the project.

Our framework addresses these gaps by supporting early-stage discovery-related activities, such as problem identification, KPI definition, feasibility assessment, and stakeholder alignment, to enable more informed and quality-driven AI innovation projects. Hence, we emphasize that the unique aspect of our contribution is a proposal that brings together aspects of an AI project that can be considered from the project's discovery stage and will assist in decision-making so that the project has higher quality and meets stakeholders' expectations.

\section[\MakeLowercase{Background}]{Background} 


\subsection{ISO/IEC 12207:2017 and ISO/IEC 5338:2023}

ISO/IEC/IEEE 12207:2017 \cite{iso12207}, entitled “Systems and Software Engineering - Software Life Cycle Processes”, is an international standard that defines a process for software engineering (SE) throughout the life cycle of a software. It is jointly published by the International Organization for Standardization (ISO), the International Electrotechnical Commission (IEC), and the Institute of Electrical and Electronics Engineers (IEEE). The standard is widely used across industries to ensure that complex systems are efficiently and effectively designed, developed, operated, and maintained.

The standard covers all phases of the software life cycle and defines a set of processes and activities necessary for SE. These processes include i) agreement; ii) project organization; iii) technical design; and iv) technical
support. It is designed to be compatible with and complementary to other standards, such as ISO/IEC/IEEE 15288:2023, which focuses on systems engineering, and ISO/IEC 5338:2023 [25], which introduces AI-specific processes and updates requirements engineering practices for AI systems.

Next, we present key points about ISO/IEC/IEEE 12207:2017, essential for the context of this proposal:

\begin{itemize}
    \item \textit{Requirements Engineering}: emphasizes the importance of identifying and documenting software requirements to guide software design and development, with dedicated processes for requirement elicitation and specification.
    \item \textit{Risk Management}: addresses risk identification and mitigation, along with quality assurance, throughout the development process to ensure software reliability.
    \item \textit{Adaptability}: offers flexibility to suit different types of software systems and organizational contexts, making it suitable for innovative and AI-driven projects.
    \item \textit{Stakeholder involvement}: highlights the need for continuous engagement with stakeholders to ensure their expectations are met throughout the life cycle.
    \item \textit{Specification and Traceability}: stresses the importance of thorough documentation and regular reviews to maintain alignment with requirements and ensure expected functionality.
\end{itemize}

In this work, we adopt this standard to support the definition of the requirements engineering process and associated artifacts, aiming to ensure problem identification, quality, and traceability in the development of AI-based systems.



\subsection{Design Thinking}


Some works highlight the use of Design Thinking (DT) and RE \cite{HehnUseDT,alhazmiIntegratingDT,HehnIntegratingDT,conteBenefitsChallengesDT,HehnDT4RE}. It is a structured approach to problem-solving and has been used to develop innovative products that explores needs and integrates an agile and flexible environment to solve complex and ill-defined problems. 

After examining the two approaches, we identified that ISO/IEC 12207 and DT can complement each other in proposing support for the discovery of AI innovation projects. DT follows a structured process represented by the Double Diamond model, which alternates between divergent and convergent thinking phases—first to explore and define the problem, then to ideate and prototype solutions.\cite{AnalysingDoubleDiamondDTGustafsson2019}. The focus of our proposal is on the first diamond to support problem identification. Identifying problems and detailing them is challenging, as is establishing communication with stakeholders. Other contributions already support RE ideation~\cite{alonso2025define} and specification activities~\cite{villamizar2024identifying}.

\section{Methodological Path}
\label{sec:solution}

Our proposal was conceived and refined based on the Technology Transfer Model (TTM) \cite{pfleeger}, which supports the development of artifacts (e.g., models, methods, and processes) to address specific challenges.

\begin{figure}[h]
\centering
\includegraphics[width=8 cm]{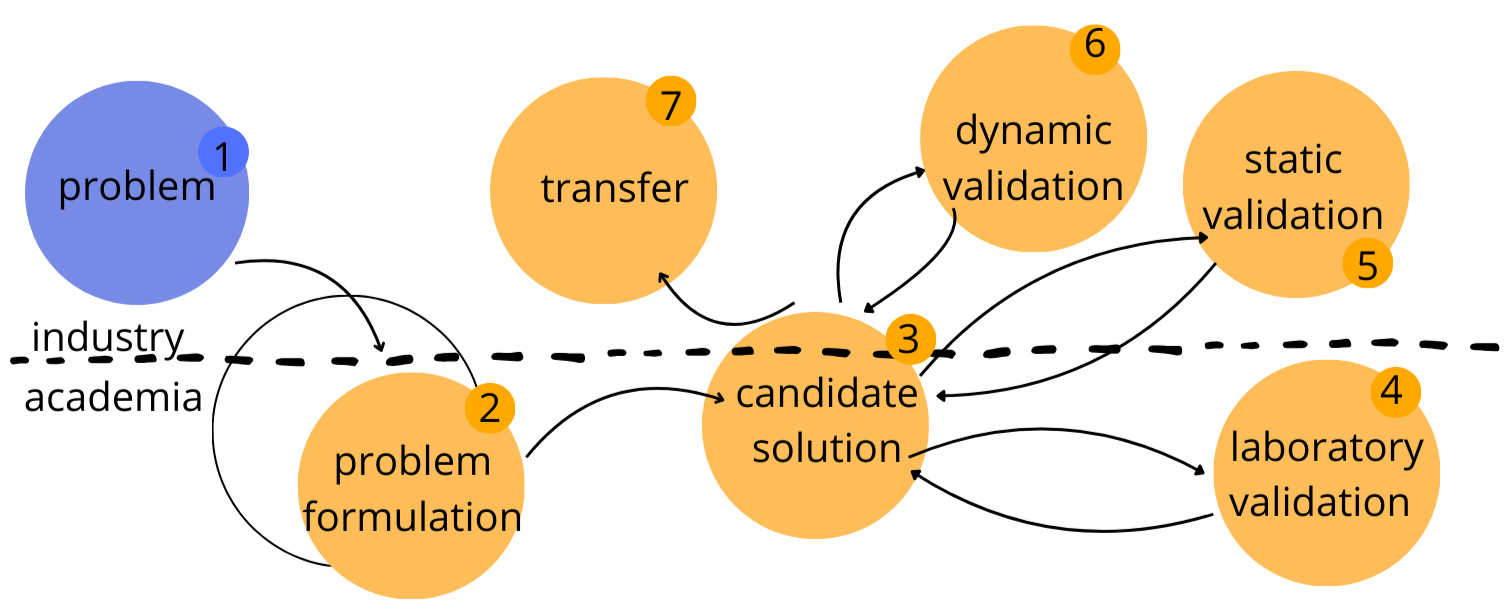}
\caption{Transfer Technology Model adapted by \cite{pfleeger}.}
\label{fig:mtt}
\end{figure}

As illustrated in Figure \ref{fig:mtt}, TTM begins with the identification of a real-world industry need, supported by findings in the literature. Based on this need, a candidate solution is designed and iteratively evaluated in various scenarios. Each evaluation cycle informs refinements to the artifact, enhancing its relevance and value for both research and practice. The next sections detail how the TTM was applied to conceive DIP-AI.

\subsection{Problem identification}

We conducted interviews with coordinators of several of AI and innovation projects. We discovered the difficulty in understanding the underlying problem associated with data-driven tasks, ensuring effective communication among stakeholders, and aligning domain-specific and technical knowledge\cite{Martinscibse}. These challenges were also found by Alves et al. \cite{alvesAPKalinowski-statusquo}.


\subsection{Problem formulation}

We conducted a tertiary study, that is, a review of secondary studies \cite{MartinsJSERD}. We searched for studies that related RE and AI in databases such as ACM DL, IEEE Xplore, Scopus, Engineering Village, Wiley, and Web of Science. We obtained 1280 studies and included 9 for full reading and data extraction. We identified gaps in RE for AI systems, particularly in the early stages of problem and data identification, communication and collaboration with stakeholders, metrics and acceptance criteria identification, NFR and trade-offs analysis, and traceability.


Given this context, we formulated the following research question: \textbf{"How can we support problem discovery to enhance the effectiveness of AI innovation projects?"}. In response, we proposed a candidate solution described in the following section.

\subsection{Candidate Solution: DIP-AI}

\subsubsection{DIP-AI Construction Process}

We studied the RE-related ISOs technical processes. We listed the activities and tasks of each process and highlighted what we considered essential for executing the RE phase so that they would not be suppressed. We also defined activities with experienced requirements engineers and the support of a data scientist. The activities were discussed iteratively over four weeks. 

Regarding the \textit{business analysis} process, we included the following activities: (i) define the scope of the problem and analyze complaints, (ii) characterize the solution space and identify alternative solutions, (iii) evaluate alternative solutions, and (iv) maintain traceability. For the \textit{stakeholders' needs definition} process, we defined the following activities: (i) identify stakeholders; (ii) define needs, context of use, and scenarios; (iii) classify and prioritize needs; (iv) identify restrictions and relationships with non-functional requirements; (v) analyze the set of requirements and define performance measures; (vi) discuss and give feedback; and (vii) obtain agreement and maintain traceability.

Concerning the \textit{system requirements definition} process, we chose the activities as follows: (i) define the functional boundary; (ii) identify modes of operation, implementation restrictions, and risks, and define requirements; (iii) analyze requirements; and (iv) obtain agreement and maintain traceability. Finally, we defined the following activities for the \textit{system analysis} process: (i) identify contexts and assumptions, analyze the results, establish conclusions and recommendations, record results, and (ii) maintain traceability. We believe that a simplified guide process can aid experimental and iterative AI and innovation development. 

After defining the essential activities, we aligned them with the DT Double Diamond model. In the empathize phase, we positioned activities from the business analysis process. The define phase encompasses tasks related to the stakeholder needs definition process. The ideate phase integrates activities from the system requirements definition process. The system analysis process spans across all three phases, supporting iterative refinement and stakeholder alignment. Finally, the prototype phase involves the development, evaluation, and refinement of ML-enabled prototypes until a solution that meets stakeholder expectations is achieved \cite{MartinsWER}.

\begin{table*}[]
\centering
\caption{Relation between ISO/IEC/IEEE 12207 modified by ISO/IEC/IEEE 5338 processes and DIP-AI Canvas}
\resizebox{\textwidth}{!}{%
\begin{tabular}{|l|l|l|l|l|l|}
\hline
\textbf{Process/Fields Goups} & \textbf{\begin{tabular}[c]{@{}l@{}}Question about\\  context projects\end{tabular}} & \textbf{5W2H} & \textbf{GUT Matrix} & \textbf{ISO 31000} & \textbf{Fesibility Analysis} \\ \hline
\textbf{\begin{tabular}[c]{@{}l@{}}Business or mission\\  analysis process\end{tabular}} & \begin{tabular}[c]{@{}l@{}}Define scope of the problem and \\ characterize the solution space\end{tabular} &  &  &  & \begin{tabular}[c]{@{}l@{}}Evaluate classes of \\ alternative solutions\end{tabular} \\ \hline
\textbf{\begin{tabular}[c]{@{}l@{}}Stakeholder needs and \\ requirements definition process\end{tabular}} &  & Define context of use & \begin{tabular}[c]{@{}l@{}}Prioritize and \\ classify needs\end{tabular} & Identify constraints &  \\ \hline
\textbf{\begin{tabular}[c]{@{}l@{}}System requirement \\ definition process\end{tabular}} &  & \begin{tabular}[c]{@{}l@{}}Identify and define \\ needs and justifications\end{tabular} & Prioritize needs & \begin{tabular}[c]{@{}l@{}}Identify risk-related \\ requirements\end{tabular} &  \\ \hline
\textbf{System analysis process} &  &  &  &  & \begin{tabular}[c]{@{}l@{}}Define strategy and \\ perform analysis\end{tabular} \\ \hline
\end{tabular}%
}
\label{tab:relationISO}
\end{table*}



Each DT phase also has expected outcomes aligned with ISO standards. The empathize phase aims to uncover user and business needs. The definition phase results in the articulation of challenges, problems, and evaluation criteria. 

For each phase, we specify relevant information and subprocesses suitable for AI innovation projects. Based on this process, we formulated guiding questions to assist in problem identification, solution analysis, and data-driven stakeholder engagement critical to this development context. Table \ref{tab:relationISO} presents the relationship between the main constructs of the canvas and ISO 12207 processes.

The 5W2H management tool has been used to make the planning and execution of projects, tasks, and actions clearer, ensuring that all relevant information is provided, assigned, and understood by all involved \cite{gallegos}. We identified that to understand the problem we needed to answer some basic questions such as: \textbf{What} are the businesses involved? \textbf{How} are they carried out? \textbf{Who} are the stakeholders? \textbf{Where} do the processes take place in the organization? \textbf{How} do these processes happen and \textbf{why}? \textbf{What} are the costs and impacts of these activities? Aligned with the 5W2H questions. These questions help better characterize the organization's existing processes and problems. This allows for greater insight into the project management landscape and better alignment with stakeholders.

The GUT Matrix prioritizes problems based on Gravity, Urgency, and Trend, helping guide decision-making in dynamic project contexts \cite{gallegos}. This GUT matrix classifies problems to establish ordered priorities to deal with problems efficiently. Therefore, the processes, problems, and tasks are prioritized (using the GUT matrix). Then the feasibility analysis is carried out by collecting information such as input and output data, algorithms, metrics, references, etc.

\begin{figure*}[h]
\centering
\includegraphics[width=\textwidth]{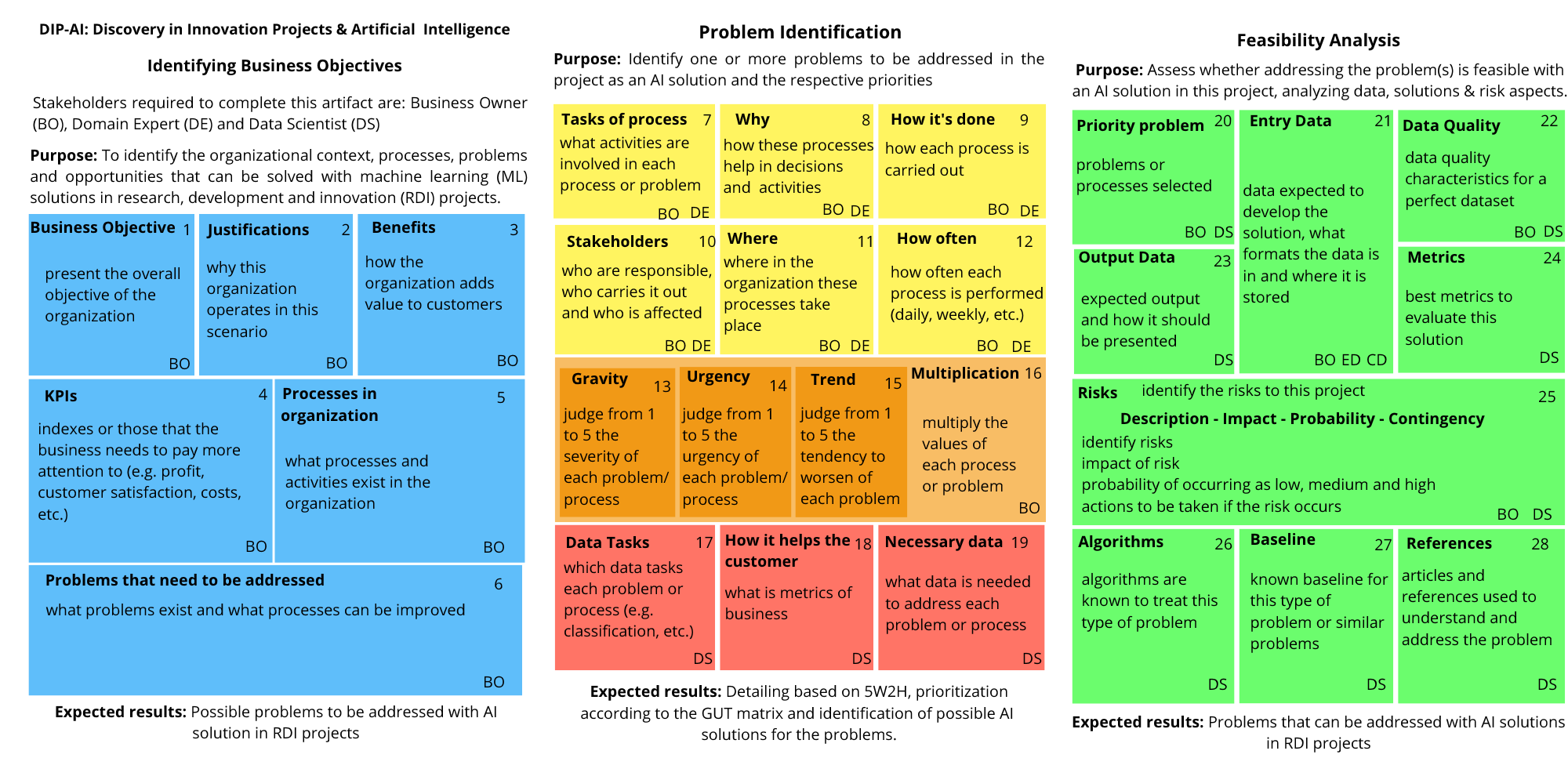}
\caption{DIP-AI Canvas.}
\label{fig:dip-ai}
\end{figure*}

We also identified the importance of understanding the task data associated with the problem as part of feasibility, as this determines whether an AI solution is applicable to the project. We refined DIP-AI by discussing it with project coordinators. The current version of the DIP-AI framework canvas is presented in Figure \ref{fig:dip-ai} and details about its structure follow.

\subsubsection{DIP-AI Framework Canvas}

The proposed canvas is structured into three parts, following the first diamond of the Design Thinking model. The first part corresponds to the empathy and immersion phase, aimed at understanding the client's context and pain points. The second part focuses on defining and prioritizing problems and project objectives. The third part refers to ideation, where possible solutions for the selected priority problem are outlined, followed by a feasibility analysis.

Stakeholder involvement—an acknowledged challenge in AI projects~\cite{MartinsWER}—is central to using the canvas properly. In the first phase, fields are completed primarily by the Business Owner (BO), who provides context, organizational background, existing challenges, and identifies other relevant stakeholders. In the second phase, the Domain Expert (DE) joins the BO to characterize and prioritize organizational processes, problems, gaps, and opportunities. At the end of this phase, the Data Scientist (DS) identifies which of the prioritized elements correspond to data-related tasks. In the third phase, the DS answers more advanced questions to evaluate the feasibility of developing AI-based solutions for the selected tasks.

By guiding this multi-stakeholder collaboration and addressing specific AI-related challenges, the canvas offers a practical approach tailored for AI innovation projects \cite{Martinscibse,alvesAPKalinowski-statusquo}. It was designed to promote an iterative and adaptable approach suited for AI innovation projects, aimed at facilitating communication through its visual structure, enabling the alignment of business objectives and technical feasibility with data, and supporting the understanding and prioritization of problems and processes.

\subsection{Validation}

The objective of these stages is to progressively refine the candidate solution to ensure its suitability for industry adoption. Following the Technology Transfer Model (TTM), \textbf{we carried out three evaluations in distinct contexts:} (i) an academic validation with graduate students in Software Engineering for AI, who used the approach to identify problems in their final course projects; (ii) a static validation in an industrial setting, where RDI professionals retrospectively applied the approach to real project artifacts—this type of validation is conducted offline, with real practitioners and data, but outside the context of active development; and (iii) a dynamic validation, embedded in a real-time industrial case study within an ongoing project. These iterative assessments were designed to promote early issue identification, enhance user satisfaction, foster continuous improvement, and build trust in the solution’s applicability. 
An approved ethics review was conducted prior to performing validations involving human participants. Subsequently, a pilot study was carried out with five graduate students to assess the validation procedure and data analysis methods, as well as to identify potential issues \footnote{CAAE: 81475724.4.0000.5083}.

\subsection{Release the solution}

In the future, we plan to implement the proposed improvements and develop documentation to effectively transfer the technology.

\section{Study Case Design}
\label{sec:study}
This investigation is structured as an evaluative case study involving participants actively engaged in a discovery phase to identify and characterize a viable problem for the project. The methodological approach was guided by the case study research principles outlined by Runeson \textit{et al.} \cite{Runeson2008}.

\subsection{Goal and Research Questions}

We followed the Goal-Question-Metric (GQM) goal definition template \cite{Basili1988} to define our research goal as follows: \textbf{\textit{Analyze} DIP-AI with the \textit{purpose} of characterization \textit{with respect to} the problem discovery capability, overall acceptance, and criticisms and suggestions for improvement, from the point of view of participants in the context of IAC AI innovation projets.}

Based on this objective, we formulated the following Research Questions (RQs):
\begin{itemize}
    \item \textbf{RQ1: How well does DIP-AI contribute to problem discovery in AI innovation projects?} To answer this question, we closely followed the experience of applying DIP-AI, analyzed the completed DIP-AI canvas artifact, specific follow-up questions, and considered the concrete result of the project (which, at the time of writing, has been completed and deployed at the industrial partner).
    \item \textbf{RQ2: How well is DIP-AI accepted in practice?} To answer this question, the follow-up questionnaire was designed based on the Technology Acceptance Model (TAM 3)~\cite{tam3}.
    \item \textbf{RQ3: How can DIP-AI be further improved?} To answer this question, we qualitatively analyzed open-text responses to the follow-up questionnaire regarding criticisms and improvement suggestions.
\end{itemize}

\subsection{Validation Context}

The context of this study is an industry-academia collaboration innovation project initiative designed to enhance the education of undergraduate and postgraduate students. In this initiative, participants play a central role in the design and development of innovative AI-based technology solutions aimed at addressing concrete business challenges of real industry partners \cite{romao2024sbes}. 
~The industrial partner involved in the particular case study is a technology company that develops digital solutions for multiple clients, with a product portfolio focused on technological security across various industry sectors. 

Problem discovery is an important part of an AI innovation project, enabling the identification of problems rooted in the genuine needs of the industrial partner. The project in question was suitable for our case study as it did not yet have a clearly defined problem. We spent a month on-site accompanying the participants whilecarrying out the Discovery phase of the project with the support of DIP-AI. We participated in the onboarding meeting with the customer, daily meetings, and applied the questionnaire.

\begin{figure}[h]
\centering
\includegraphics[width=8 cm]{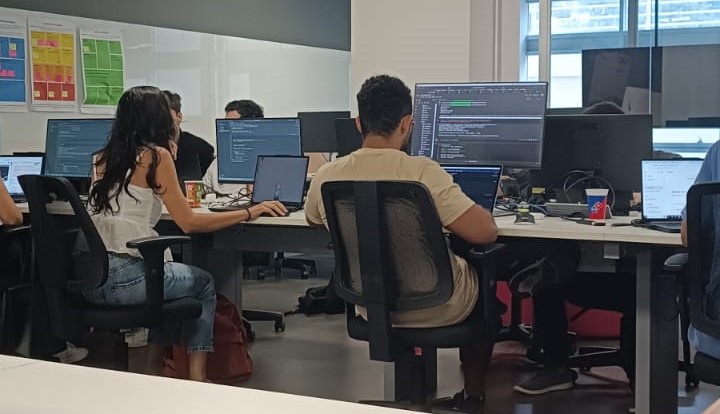}
\caption{Participants in the Case Study Context and the Use of DIP-AI in the background.}
\label{fig:participants}
\end{figure}

Regarding the profile of the participants, the project team had eight undergraduate students and two doctoral students. All participants had at least one year of prior experience in AI innovation projects, having been involved in between two and six such projects. Six of the participants declared themselves experienced and skilled in identifying problems in AI projects. More details about the participants' profiles are shared with other materials from this evaluation~\footnote{\url{https://zenodo.org/records/15732851}}. Figure \ref{fig:participants} illustrates the participants in the context of the case study and the DIP-AI framework used in the background. 

\begin{figure}[h]
\centering
\includegraphics[width=8 cm]{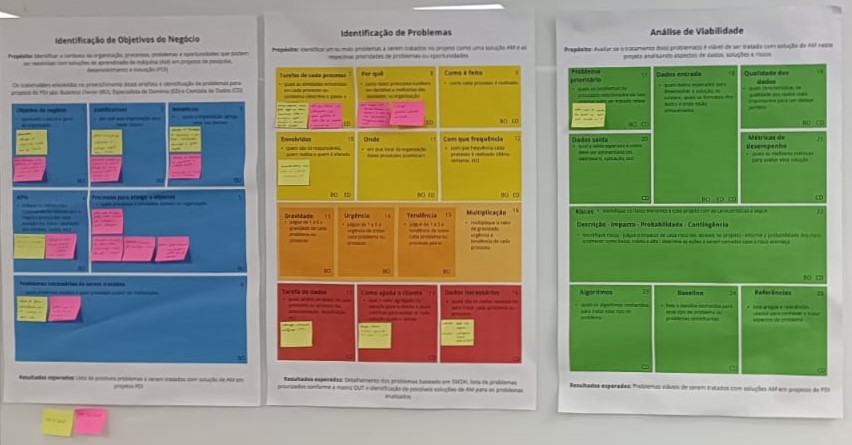}
\caption{DIP-AI completed by participants in the case study.}
\label{fig:DIP-AI-complete}
\end{figure}

Figure \ref{fig:DIP-AI-complete} illustrates the DIP-AI framework canvas completed during the case study. The canvas was first introduced to the project coordinators on January 15, 2025, when the constructs employed in the canvas, the potential benefits of its evaluation, and the components of the artifact were presented. On January 22, 2025, the first author conducted a presentation to the participants, providing an overview of the canvas and explaining each field in detail. During this session, participants posed questions about its use and engaged in discussions about the relevance of the canvas and how it might help to mitigate challenges in their project. Each presentation lasted approximately one hour.

In the subsequent days, the first author documented difficulties reported by participants in their respective project that could potentially be addressed through the proposed approach. Additional contextual information about the industrial partner and the project challenge was also gathered. On January 29, 2025, the project objective was revised in collaboration with the partner company and the innovation unit due to feasibility concerns. As a result, the canvas was completed for a second time on January 30, 2025, between 3:30 p.m. and 6:00 p.m. During this session, participants identified and prioritized multiple project-related tasks using the GUT Matrix.

In the days that followed, participants continued to raise questions about specific canvas fields, although they did not resume completion immediately. The process was resumed on February 6, 2025, from 3:20 p.m. to 4:45 p.m., at which point participants were better prepared to engage with the artifact. Having met with company representatives to clarify uncertainties and deepened their understanding of the problem, they completed all remaining fields.


\subsection{Data Collection}

Data collected throughout the experience of the participants applying DIP-AI involved the completed DIP-AI canvas artifact, the follow-up questionnaire, and information on the concrete result of the project.

\begin{figure*}[]
\centering
\includegraphics[width=\textwidth]{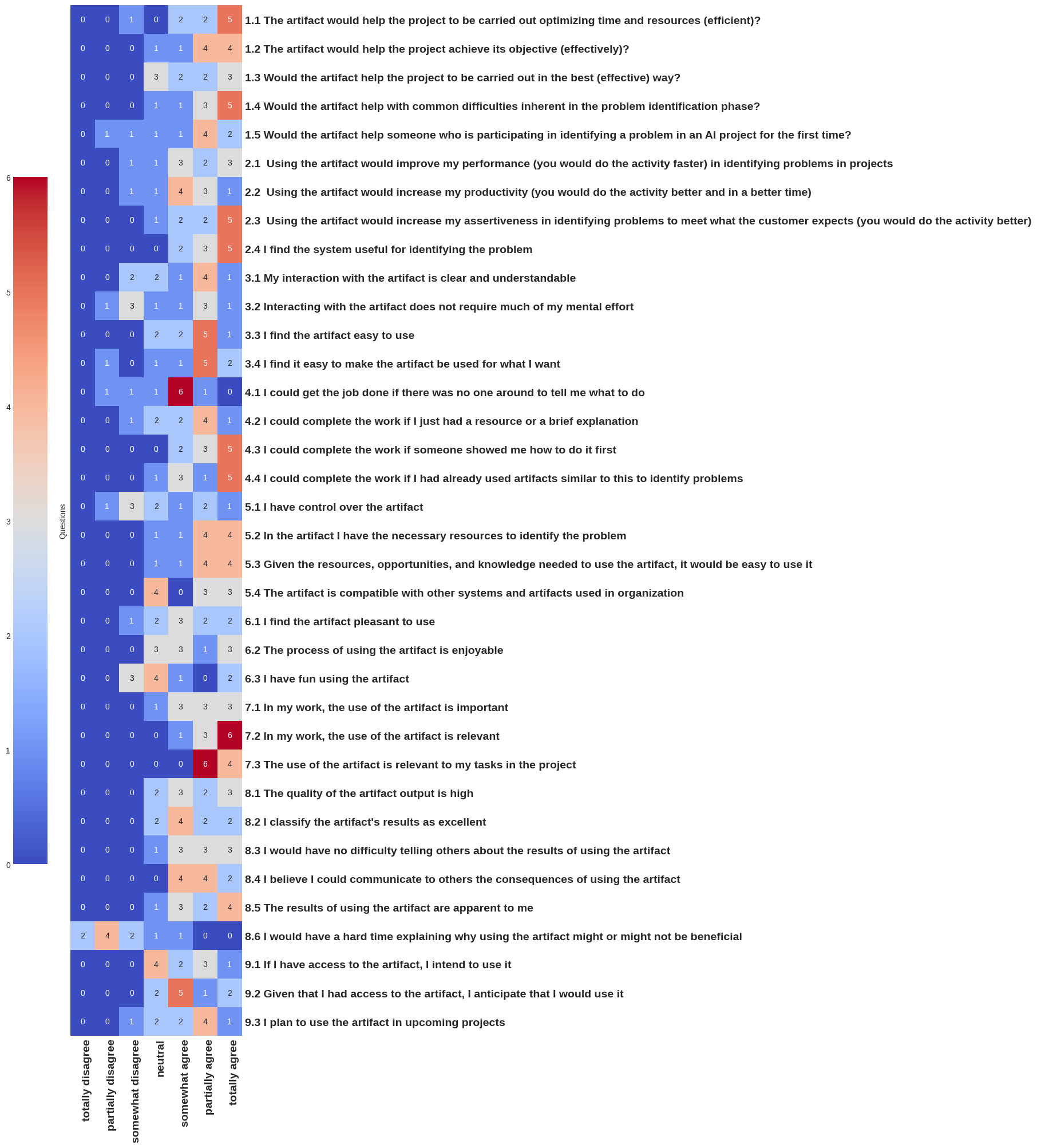}
\caption{Heat Map of Participant's Perception (TAM 3-based)}
\label{fig:heatMap}
\end{figure*}

For RQ1 (contributions to discovery) we considered the overall problem discovery experience, questions about perceptions of contributions of the artifact (questions that we created and that are shown in group 1 in Figure \ref{fig:heatMap}), and the open question "In the projects in which you participated, what were the difficulties in identifying the problem or machine learning task to be addressed in that project? Could they be mitigated by using this artifact?''.

For RQ2 (acceptance), we considered the answers to the TAM-based~\cite{tam3} questions. Considering the research objectives and context, the study focused on Technology Acceptance Model (TAM) constructs related to perceived usefulness—specifically job relevance, output quality, and result demonstrability—and perceived ease of use, including computer self-efficacy, perception of external control, perceived enjoyment, and objective usability. To ensure methodological rigor, the original semantics of the selected constructs were preserved, and internal reliability was assessed using Cronbach’s Alpha, yielding a value of 0.95, which indicates excellent reliability. We selected eight of the TAM 3 constructs (questions of groups 2 to 9 in Figure \ref{fig:heatMap}), which are: usefulness, ease of use, ability to perform the task using the artifact, control of the artifact to perform the task, pleasure in using the artifact, perceived relevance of the artifact, output quality and demonstrability, and intention to use the artifact.

Finally, for RQ3, we considered the answers to the following open questions:

\begin{itemize}
\item  What limitations, recommendations, or opportunities for improvement do you identify in this artifact?
\item Do you consider that any questions or fields are missing from the artifact?
\item Do you find the current structure and organization of the artifact to be clear and appropriate?
\end{itemize}

\subsection{Analysis Procedures}

For the quantitative analysis of the TAM questions, we performed a reliability analysis using Cronbach's Alpha. Reliability indicates how reliably or accurately a questionnaire or test measures a true value. Cronbach's alpha is given by \( \alpha = \frac{N \bar{c}}{\bar{v} + (N - 1)\bar{c}} \), where \( N \) represents the number of items, \( \bar{c} \) is the average inter-item covariance, and \( \bar{v} \) is the average variance. The quantitative analysis was conducted by aggregating the number of responses for each questionnaire item, based on a seven-point Likert scale to assess the perceived contribution of the framework.

The qualitative analysis of open questions was conducted using the Atlas.TI tool \cite{atlas}, applying open and axial coding procedures from Grounded Theory \cite{corbinStrauss2014basics}.The method we used was to gather participants' responses to each open-ended question in the questionnaire. We then analyzed the contributions, coded each response, grouped the responses, and established possible relationships. Since we were unable to reach data saturation due to the small sample of participants, we did not perform selective coding.

\section{Results}

In Figure \ref{fig:heatMap}, we present a heat map that represents the responses of the 10 participants to the 36 Likert scale items (the five created ones for artifact contribution, and the 31 from the 8 selected TAM 3 constructs) that form the perception questionnaire of this research. The questions are numbered in the form X.Y where X represents the group of questions and Y enumerates the questions in a group. The values in each field of the heat map represent the number of responses to a given question. We obtained a Cronbach's Alpha of 0.950, which means that the internal consistency of the questionnaire can be considered excellent. Hereafter, we answer each of the research questions.

\subsection{RQ1: DIP-AI's contribution to problem discovery}


Regarding RQ1, considering the overall discovery experience, the team filled the DIP-AI canvas and used it to guide the identification of the problem. In fact, it supported a strategic redefinition of the problem by analyzing risks and opportunities early. The team started working on understanding a particular problem, but then identified risks and a more promising alternative, changing direction. It is noteworthy that at the time of writing, an MVP solving the identified problem was successfully developed and deployed at the industry partner.

Furthermore, based on the first group of questions, it is possible to observe that participants reported that use of the DIP-AI canvas positively influenced project execution by optimizing resources and supporting the achievement of objectives, particularly because problem identification was conducted accurately, efficiently, and effectively.

Additionally, participants acknowledged that the use of the DIP-AI artifact helped address common challenges encountered during the problem discovery phase in innovation and AI-related projects. However, two participants disagreed about the artifact's usefulness for individuals identifying a problem in a project for the first time.

Participants' answers to the question "In the projects in which you participated, what were the difficulties in identifying the problem or machine learning task to be addressed in that project? Could they be mitigated by using this artifact?''.

\begin{itemize}
    \item P3: "I believe that using this artifact right from the beginning could help us look at data and request possible pending issues that may exist.''
    \item P6: "I believe that using this artifact the difficulties could be mitigated''
    \item P8: "Lack of clarity of processes related to the problem and there is a need to structure the necessary data in a clear way.''
    \item P7: "Lack of clarity in communication with the customer and in the company's processes are difficulties that could be mitigated, since the artifact would help the team prioritize elements of an AI task.''
\end{itemize}

\begin{mdframed}
\noindent
\textbf{RQ1 Answer:} The use of DIP-AI was suitable, and it was considered to contribute positively to the discovery phase of the AI innovation project.
\end{mdframed}

\subsection{RQ2: DIP-AI overall acceptance}

Participants expressed positive perceptions regarding the \textbf{usefulness} of the artifact (as reflected in the responses from Group 2). They considered the artifact to be beneficial in several ways: it enhances productivity, increases assertiveness in decision-making, and supports effective problem identification. 

Regarding \textbf{ease of use} (Group 3), four of the participants reported that the artifact requires a lot of mental effort. And only one student reported that he did not find it easy to use the artifact for what he wanted. 

Regarding the group of questions (Group 4) related to the \textbf{ability to identify problems} using the artifact, two participants reported that they would have difficulty completing the activity without guidance. Additionally, one participant indicated that relying solely on a support resource or a brief explanation would not be sufficient to carry out the task effectively. These perceptions align with comments from other participants who noted that the artifact is not entirely intuitive or easy to use.

Regarding the perception of \textbf{controlling the artifact} (Group 5) to perform the activity, three participants disagreed with having control over the artifact. Perhaps because they consider it not a very customizable tool. 

Regarding the dimension of \textbf{pleasure} (Group 6) in performing the activity with the artifact, one participant reported not experiencing any sense of enjoyment, and three others stated that they did not find the task engaging or fun. These responses are consistent with the broader perception that the artifact is not particularly easy to use, as it is associated with a cognitively demanding activity that requires significant mental effort.

No negative perceptions were reported regarding the \textbf{relevance} of the artifact (Group 7) or the importance of its use in supporting tasks related to AI project development. Similarly, there were no negative comments concerning the demonstrability and quality of the artifact’s results (Group 8). On the contrary, participants emphasized that the artifact contributes positively to stakeholder communication, reinforcing its practical value in project contexts.

Finally, nine responses were positive regarding the \textbf{intention to use} (Group 9) the artifact in the future.

\vspace{0.2cm}

\begin{mdframed}
\noindent
\textbf{RQ2 Answer:} The use of DIP-AI was well accepted by agile team involved in innovation project.
\end{mdframed} 

\subsection{RQ3: DIP-AI criticisms and suggestions}

Participants generally reported that the artifact was well-structured and adequately organized, with no missing fields. They also mentioned that the use of the framework could help mitigate challenges previously encountered in other projects.

Participants indicated that DIP-AI is particularly valuable during the discovery phase of AI innovation projects. Three responses highlighted that the framework contributed to understanding the problem and facilitated communication with stakeholders. Additionally, four participants noted that the DIP-AI canvas played a key role in stakeholder discussions concerning the availability of data, an essential aspect for the success of such projects. Participants also identified specific advantages of using the artifacts: one response emphasized the artifact’s role in task prioritization, and three others mentioned that the canvas helped reduce common difficulties in innovation and AI initiatives.

Despite the positive feedback, participants also offered several suggestions for improvement. One participant noted that the artifact is extensive and proposed completing it in iterative rounds. Another suggested that some fields could be merged to reduce the overall complexity. Additional suggestions included the inclusion of a glossary to clarify terms and acronyms, as well as providing guidance to assist stakeholders in completing the artifact. One participant recommended enriching the "priority problem" field by linking it to problem characterization — specifically, incorporating information from the GUT matrix and task data into the analysis of findings. Another participant proposed integrating the canvas with agile methodologies, such as Scrum, to periodically reflect on the project’s objectives during each sprint. Finally, two respondents suggested that the canvas would benefit from tool support to enhance user interaction and usability.

\begin{itemize}
    \item P3: "I believe that using this artifact right from the beginning could help us look at data and request possible pending issues that may exist.''
    \item P6: "I would add, along with the algorithms, the characterization of the priority problem. For example: it is a classification problem, regression, supervised learning, unsupervised, it will involve Deep Learning, LLMs.''
    \item P8: "I would like some concepts of the artifact to be explained through a glossary, because if we did not have help, it would be confusing to understand some terms.''
    \item P7: "I would use the artifact as an aid to SCRUM, just like KANBAN. Filling out the artifact in "rounds" is one of its greatest values: giving a temporal notion of the team's perception of the project as a whole. It may even have too many fields, which could lead to exhaustion for those using the artifact and cut short the creative innovation process a bit by keeping the user too tied to a (possibly) long and closed step-by-step process. In my opinion, the artifact could benefit a lot from tool support with an inviting and clean UX.''
    \item P10: "During the dynamic, I felt that some fields were confused with each other, and perhaps, without the help of the researcher, I would not have properly understood what differentiated them. For example, the GUT matrix was a little confusing to me at first, although it was very useful. I found it more practical to use it virtually.''
\end{itemize}



\vspace{0.2cm}

\begin{mdframed}
\noindent
\textbf{RQ3 Answer:} DIP-AI contributed to problem discovery, communicating with stakeholders, and prioritization of problems. We also received suggestions to make the canvas more user-friendly, such as adding a glossary and tool support.
\end{mdframed} 

\section {Discussion} 

DIP-AI is a proposal aligned with the RE phase applied to AI projects. In AI project development processes, there is concern with the Discovery phase, where the problem is identified and characterized; the data is discovered, and the project is prepared for the other phases. Project coordinators pointed out the difficulty in identifying the problem and in building and maintaining this documentation for their projects \cite{Martinscibse}. This difficulty was also pointed in studies on the state of the art and practice \cite{villamizarMapeamento,alvesAPKalinowski-statusquo}. The approach was developed to help discovering and selecting problems to be addressed in a project. DIP-AI was useful in identifying the problem to be addressed in the project. The AI product developed and delivered within the case study was based on a WhatsApp user interface and employed LLMs for fluent communication and an ML classification model to identify potentially unsafe situations, also proactively suggesting protective actions to users. The engine is capable of interpreting multiple input formats provided through WhatsApp (QR codes, images, locations, and text messages).

In the evaluative case study, the proposed framework was perceived as both useful and relevant. The quantitative analysis supports this perception, indicating that the artifact effectively aids in problem identification. These findings are corroborated by the qualitative analysis, which reinforces the positive evaluations observed in the quantitative data. The results suggest that the framework facilitates communication and assists in collecting project-related information that may signal potential future risks, such as the viability and availability of data and models. However, perceptions regarding the ease of use were not uniformly positive. The qualitative feedback provided valuable insights on how to enhance the proposal’s usability. Suggestions include simplifying the completion process through the use of a glossary, guided assistance, or iterative completion rounds.

\section{Threats to validity}
\label{sec:threats}

In this section, we discuss the threats to the validity of our study, according to the categories presented by Runeson \textit{et al.} \cite{Runeson2008}. 

In terms of \textbf{construct validity}, potential challenges include the influence of assessment apprehension, which may mobilize participants to avoid negative feedback. To mitigate this threat, the questionnaire was administered anonymously. In addition, hypothesis guessing emerges as a potential threat, where participants may have provided overly positive responses, anticipating the research being conducted. To minimize this problem, we clarified that the questionnaire would be used to investigate ways to understand and improve the experience of identifying problems in projects. Participants were encouraged to be honest and critical in order to improve the approach in the future. 

For \textbf{internal validity}, issues such as diffusion or imitation of treatments come to the fore. Some participants may have sought information about the motivation for some responses from their peers, compromising the integrity of our study. To minimize this threat, we included open-ended questions at the end of the questionnaire to complement the quantitative analysis of the results and better understand participants’ perceptions when answering each question. Furthermore, we applied the questionnaire individually and synchronously to avoid the influence of comments among participants. 

Regarding \textbf{external validity}, we faced the selection and treatment interaction due to the non-representative sample achieved by our study. The participants chosen for the research were currently active in AI innovation projects and had already performed problem identification in previous projects. Despite this, we recognize the importance of carrying out more case study evaluations with different participant profiles and projects for generalization. 

Finally, regarding \textbf{reliability}, we conducted a single case study. However, as expected for case studies, we carefully described the context, and our case study fills a gap in the literature by identifying problems of innovation and AI projects in an IAC context with a large Brazilian security company. To improve the reliability of our analyses, we peer-reviewed all of our analysis procedures involving two independent researchers and made all data available online for audit in our open science repository.

\section{Final Remarks}
\label{sec:conclusao}

This paper presents DIP-AI, an approach developed to identify and prioritize key problems to be addressed in artificial intelligence innovation projects. The approach empowers requirements engineers — working collaboratively with business owners, domain experts, and data scientists — to uncover opportunities for applying AI to solve business problems. We introduced the DIP-AI canvas with 28 fields for identifying the business objectives, characterizing problems and opportunities, and analyzing feasibility.

We carried out a case study in the discovery phase of an ongoing project with ten team members who had previous experience with AI innovation projects. We held workshops for the participants to fill out the DIP-AI canvas, and the participants had the opportunity to ask questions about filling out the canvas and collecting information from customers.

The DIP-AI canvas was reviewed by data scientists and innovation project coordinators who provided feedback. The perceptions collected from the questionnaires were positive regarding the usefulness and relevance of the canvas. Despite this, we received negative feedback regarding the ease of use and pleasure in using the artifact. These perceptions were corroborated in the qualitative analysis, where the participants indicated that the proposal could benefit from a glossary and tool support for filling out the canvas.

The case study demonstrates that the DIP-AI canvas supports the formulation of innovative artificial intelligence solutions by aligning organizational objectives with process mapping and customer problem identification. It enhances stakeholder communication, guides the exploration process from pain point identification to feasibility assessment, and offers insights into practitioner needs, thereby informing future research aligned with the requirements engineering phases defined in ISO/IEC 12207 and ISO/IEC 5338. We also offer practical implications for industry by presenting a successful application of the framework for problem identification in a real AI innovation project developed for a large technology company.

As future work, we plan to conduct additional case studies encompassing a broader and more diverse range of artificial intelligence innovation projects. Furthermore, we intend to design and evaluate a software tool to facilitate the application of DIP-AI and to assess its usability and contributions through focus group sessions with domain experts.

 \section*{ACKNOWLEDGMENTS}

\small{We thank ExACTa, including the program sponsors, business managers, and especially the Digital Innovation team, for their support in establishing a fruitful collaboration between industry and academia. We also thank the students, mentors, and faculty enrolled in the program for their commitment to making this collaborative educational program a valuable learning experience for all involved. We thank the Brazilian Coordination for the Improvement of Higher Education Personnel (CAPES) for its financial support—Financial Code 001.}



\bibliographystyle{ACM-Reference-Format}

\begin{thebibliography}{46}


\ifx \showCODEN    \undefined \def \showCODEN     #1{\unskip}     \fi
\ifx \showDOI      \undefined \def \showDOI       #1{#1}\fi
\ifx \showISBNx    \undefined \def \showISBNx     #1{\unskip}     \fi
\ifx \showISBNxiii \undefined \def \showISBNxiii  #1{\unskip}     \fi
\ifx \showISSN     \undefined \def \showISSN      #1{\unskip}     \fi
\ifx \showLCCN     \undefined \def \showLCCN      #1{\unskip}     \fi
\ifx \shownote     \undefined \def \shownote      #1{#1}          \fi
\ifx \showarticletitle \undefined \def \showarticletitle #1{#1}   \fi
\ifx \showURL      \undefined \def \showURL       {\relax}        \fi
\providecommand\bibfield[2]{#2}
\providecommand\bibinfo[2]{#2}
\providecommand\natexlab[1]{#1}
\providecommand\showeprint[2][]{arXiv:#2}




\bibitem[Ahmad et~al\mbox{.}(2022)]%
        {ahmad2022mapeamento}
\bibfield{author}{\bibinfo{person}{Khlood Ahmad}, \bibinfo{person}{Mohamed Abdelrazek}, \bibinfo{person}{Chetan Arora}, \bibinfo{person}{Muneera Bano}, {and} \bibinfo{person}{John Grundy}.} \bibinfo{year}{2022}\natexlab{}.
\newblock \bibinfo{title}{Requirements Engineering for Artificial Intelligence Systems: A Systematic Mapping Study}.
\newblock
\newblock
\showeprint[arxiv]{2212.10693}~[cs.SE]


\bibitem[Ahmed et~al\mbox{.}(2018)]%
        {ahmed}
\bibfield{author}{\bibinfo{person}{Bakhtiyar Ahmed}, \bibinfo{person}{Thomas Dannhauser}, {and} \bibinfo{person}{Nada Philip}.} \bibinfo{year}{2018}\natexlab{}.
\newblock \showarticletitle{A Lean Design Thinking Methodology (LDTM) for Machine Learning and Modern Data Projects}. In \bibinfo{booktitle}{\emph{2018 10th Computer Science and Electronic Engineering (CEEC)}}. \bibinfo{pages}{11--14}.
\newblock
\urldef\tempurl%
\url{https://doi.org/10.1109/CEEC.2018.8674234}
\showDOI{\tempurl}


\bibitem[Alhazmi and Huang(2020)]%
        {alhazmiIntegratingDT}
\bibfield{author}{\bibinfo{person}{Alhejab Alhazmi} {and} \bibinfo{person}{Shihong Huang}.} \bibinfo{year}{2020}\natexlab{}.
\newblock \showarticletitle{{Integrating Design Thinking into Scrum Framework in the Context of Requirements Engineering Management}}. In \bibinfo{booktitle}{\emph{Proceedings of the 3rd International Conference on Computer Science and Software Engineering}}. \bibinfo{pages}{33–45}.
\newblock
\showISBNx{9781450375528}
\urldef\tempurl%
\url{https://doi.org/10.1145/3403746.3403902}
\showDOI{\tempurl}


\bibitem[Alonso et~al\mbox{.}(2025)]%
        {alonso2025define}
\bibfield{author}{\bibinfo{person}{Silvio Alonso}, \bibinfo{person}{Antonio Pedro~Santos Alves}, \bibinfo{person}{Lucas Romao}, \bibinfo{person}{H{\'e}lio Lopes}, {and} \bibinfo{person}{Marcos Kalinowski}.} \bibinfo{year}{2025}\natexlab{}.
\newblock \showarticletitle{Define-ML: An Approach to Ideate Machine Learning-Enabled Systems}. In \bibinfo{booktitle}{\emph{2025 51st Euromicro Conference on Software Engineering and Advanced Applications (SEAA)}}. Springer.
\newblock


\bibitem[Alves et~al\mbox{.}(2024)]%
        {alvesAPKalinowski-statusquo}
\bibfield{author}{\bibinfo{person}{Antonio Pedro~Santos Alves}, \bibinfo{person}{Marcos Kalinowski}, \bibinfo{person}{G{\"o}rkem Giray}, \bibinfo{person}{Daniel Mendez}, \bibinfo{person}{Niklas Lavesson}, \bibinfo{person}{Kelly Azevedo}, \bibinfo{person}{Hugo Villamizar}, \bibinfo{person}{Tatiana Escovedo}, \bibinfo{person}{Helio Lopes}, \bibinfo{person}{Stefan Biffl}, \bibinfo{person}{J{\"u}rgen Musil}, \bibinfo{person}{Michael Felderer}, \bibinfo{person}{Stefan Wagner}, \bibinfo{person}{Teresa Baldassarre}, {and} \bibinfo{person}{Tony Gorschek}.} \bibinfo{year}{2024}\natexlab{}.
\newblock \showarticletitle{Status Quo and Problems of Requirements Engineering for Machine Learning: Results from an International Survey}. In \bibinfo{booktitle}{\emph{Product-Focused Software Process Improvement}}, \bibfield{editor}{\bibinfo{person}{Regine Kadgien}, \bibinfo{person}{Andreas Jedlitschka}, \bibinfo{person}{Andrea Janes}, \bibinfo{person}{Valentina Lenarduzzi}, {and} \bibinfo{person}{Xiaozhou Li}} (Eds.). \bibinfo{publisher}{Springer Nature Switzerland}, \bibinfo{address}{Cham}, \bibinfo{pages}{159--174}.
\newblock
\showISBNx{978-3-031-49266-2}


\bibitem[Anand et~al\mbox{.}(1998)]%
        {anand98}
\bibfield{author}{\bibinfo{person}{S.~S. Anand}, \bibinfo{person}{A.~R. Patrick}, \bibinfo{person}{J.~G. Hughes}, {and} \bibinfo{person}{D.~A. Bell}.} \bibinfo{year}{1998}\natexlab{}.
\newblock \showarticletitle{A data mining methodology for cross-sales}.
\newblock \bibinfo{journal}{\emph{Know.-Based Syst.}} \bibinfo{volume}{10}, \bibinfo{number}{7} (\bibinfo{date}{May} \bibinfo{year}{1998}), \bibinfo{pages}{449–461}.
\newblock
\showISSN{0950-7051}
\urldef\tempurl%
\url{https://doi.org/10.1016/S0950-7051(98)00035-5}
\showDOI{\tempurl}


\bibitem[Atlas.TI(2024)]%
        {atlas}
Atlas.TI \bibinfo{year}{2024}\natexlab{}.
\newblock
\newblock
\newblock
\shownote{https://atlasti.com}.


\bibitem[Basili and Rombach(1988)]%
        {Basili1988}
\bibfield{author}{\bibinfo{person}{V.R. Basili} {and} \bibinfo{person}{H.D. Rombach}.} \bibinfo{year}{1988}\natexlab{}.
\newblock \showarticletitle{The TAME project: towards improvement-oriented software environments}.
\newblock \bibinfo{journal}{\emph{IEEE Transactions on Software Engineering}} \bibinfo{volume}{14}, \bibinfo{number}{6} (\bibinfo{date}{June} \bibinfo{year}{1988}), \bibinfo{pages}{758–773}.
\newblock
\showISSN{0098-5589}
\urldef\tempurl%
\url{https://doi.org/10.1109/32.6156}
\showDOI{\tempurl}


\bibitem[Cabena et~al\mbox{.}(1998)]%
        {cabena}
\bibfield{author}{\bibinfo{person}{Peter Cabena}, \bibinfo{person}{Pablo Hadjinian}, \bibinfo{person}{Rolf Stadler}, \bibinfo{person}{Jaap Verhees}, {and} \bibinfo{person}{Alessandro Zanasi}.} \bibinfo{year}{1998}\natexlab{}.
\newblock \bibinfo{booktitle}{\emph{Discovering data mining: from concept to implementation}}.
\newblock \bibinfo{publisher}{Prentice-Hall, Inc.}, \bibinfo{address}{USA}.
\newblock
\showISBNx{0137439806}


\bibitem[Corbin and Strauss(2014)]%
        {corbinStrauss2014basics}
\bibfield{author}{\bibinfo{person}{J. Corbin} {and} \bibinfo{person}{A. Strauss}.} \bibinfo{year}{2014}\natexlab{}.
\newblock \bibinfo{booktitle}{\emph{Basics of Qualitative Research: Techniques and Procedures for Developing Grounded Theory} (\bibinfo{edition}{4th} ed.)}.
\newblock \bibinfo{publisher}{SAGE Publications}.
\newblock
\showISBNx{9781483315683}
\showLCCN{2014048844}
\urldef\tempurl%
\url{https://books.google.com.br/books?id=hZ6kBQAAQBAJ}
\showURL{%
\tempurl}


\bibitem[CRISP-DM - version 2.0(2008)]%
        {CRISP-DM}
CRISP-DM - version 2.0 \bibinfo{year}{2008}\natexlab{}.
\newblock \bibinfo{title}{CRISP-DM - CRoss-Industry Standard Process for DM}.
\newblock
\newblock
\newblock
\shownote{(version 1.0 of 1999)}.


\bibitem[Dewalt(2017)]%
        {dewalt}
\bibfield{author}{\bibinfo{person}{Kevin Dewalt}.} \bibinfo{year}{2017}\natexlab{}.
\newblock \bibinfo{title}{The AI Canvas - The strategic framework for enterprise deep learning}.
\newblock
\newblock
\urldef\tempurl%
\url{https://medium.com/the-business-of-ai/the-ai-canvas-7a8717cddbe9}
\showURL{%
\tempurl}


\bibitem[Dorard(2015)]%
        {dorard2015}
\bibfield{author}{\bibinfo{person}{Louis Dorard}.} \bibinfo{year}{2015}\natexlab{}.
\newblock \bibinfo{title}{Machine Learning Canvas}.
\newblock
\newblock
\urldef\tempurl%
\url{https://www.machinelearningcanvas.com/}
\showURL{%
\tempurl}


\bibitem[Fayyad et~al\mbox{.}(1996)]%
        {fayyad}
\bibfield{author}{\bibinfo{person}{Usama Fayyad}, \bibinfo{person}{Gregory Piatetsky-Shapiro}, {and} \bibinfo{person}{Padhraic Smyth}.} \bibinfo{year}{1996}\natexlab{}.
\newblock \showarticletitle{Knowledge discovery and data mining: towards a unifying framework}. In \bibinfo{booktitle}{\emph{Proceedings of the Second International Conference on Knowledge Discovery and Data Mining}} (Portland, Oregon) \emph{(\bibinfo{series}{KDD'96})}. \bibinfo{publisher}{AAAI Press}, \bibinfo{pages}{82–88}.
\newblock


\bibitem[Gallegos(2023)]%
        {gallegos}
\bibfield{author}{\bibinfo{person}{Raphael Augusto~Parreiras Gallegos}.} \bibinfo{year}{2023}\natexlab{}.
\newblock \bibinfo{booktitle}{\emph{Ferramentas de Gestão Voltadas para Melhoria da Qualidade nas Empresas}}.
\newblock \bibinfo{publisher}{Freitas Bastos Editora}.
\newblock


\bibitem[Giray(2021)]%
        {giray2021software}
\bibfield{author}{\bibinfo{person}{G{\"o}rkem Giray}.} \bibinfo{year}{2021}\natexlab{}.
\newblock \showarticletitle{A software engineering perspective on engineering machine learning systems: State of the art and challenges}.
\newblock \bibinfo{journal}{\emph{Journal of Systems and Software}}  \bibinfo{volume}{180} (\bibinfo{year}{2021}), \bibinfo{pages}{111031}.
\newblock


\bibitem[Goodell et~al\mbox{.}(2021)]%
        {goodell2021artificial}
\bibfield{author}{\bibinfo{person}{John~W Goodell}, \bibinfo{person}{Satish Kumar}, \bibinfo{person}{Weng~Marc Lim}, {and} \bibinfo{person}{Debidutta Pattnaik}.} \bibinfo{year}{2021}\natexlab{}.
\newblock \showarticletitle{Artificial intelligence and machine learning in finance: Identifying foundations, themes, and research clusters from bibliometric analysis}.
\newblock \bibinfo{journal}{\emph{Journal of Behavioral and Experimental Finance}}  \bibinfo{volume}{32} (\bibinfo{year}{2021}), \bibinfo{pages}{100577}.
\newblock


\bibitem[Gustafsson(2019)]%
        {AnalysingDoubleDiamondDTGustafsson2019}
\bibfield{author}{\bibinfo{person}{Daniel Gustafsson}.} \bibinfo{year}{2019}\natexlab{}.
\newblock \showarticletitle{Analysing the Double diamond design process through research \& implementation}.
\newblock
\urldef\tempurl%
\url{https://api.semanticscholar.org/CorpusID:199016905}
\showURL{%
\tempurl}


\bibitem[Habibullah et~al\mbox{.}(2023)]%
        {habibullah2023}
\bibfield{author}{\bibinfo{person}{Khan~Mohammad Habibullah}, \bibinfo{person}{Gregory Gay}, {and} \bibinfo{person}{Jennifer Horkoff}.} \bibinfo{year}{2023}\natexlab{}.
\newblock \showarticletitle{Non-functional requirements for machine learning: understanding current use and challenges among practitioners}.
\newblock \bibinfo{journal}{\emph{Requirements Engineering}} \bibinfo{volume}{28}, \bibinfo{number}{2} (\bibinfo{date}{Jan.} \bibinfo{year}{2023}), \bibinfo{pages}{283–316}.
\newblock
\showISSN{1432-010X}
\urldef\tempurl%
\url{https://doi.org/10.1007/s00766-022-00395-3}
\showDOI{\tempurl}


\bibitem[Hehn et~al\mbox{.}(2020)]%
        {HehnIntegratingDT}
\bibfield{author}{\bibinfo{person}{Jennifer Hehn} {et~al\mbox{.}}} \bibinfo{year}{2020}\natexlab{}.
\newblock \showarticletitle{{On Integrating Design Thinking for Human-Centered Requirements Engineering}}.
\newblock \bibinfo{journal}{\emph{IEEE Software}} \bibinfo{volume}{37}, \bibinfo{number}{2} (\bibinfo{year}{2020}), \bibinfo{pages}{25--31}.
\newblock
\urldef\tempurl%
\url{https://doi.org/10.1109/MS.2019.2957715}
\showDOI{\tempurl}


\bibitem[Hehn and Uebernickel(2018)]%
        {HehnUseDT}
\bibfield{author}{\bibinfo{person}{Jennifer Hehn} {and} \bibinfo{person}{Falk Uebernickel}.} \bibinfo{year}{2018}\natexlab{}.
\newblock \showarticletitle{{The Use of Design Thinking for Requirements Engineering: An Ongoing Case Study in the Field of Innovative Software-Intensive Systems}}. In \bibinfo{booktitle}{\emph{2018 IEEE 26th International Requirements Engineering Conference (RE)}}. \bibinfo{pages}{400--405}.
\newblock
\urldef\tempurl%
\url{https://doi.org/10.1109/RE.2018.00-18}
\showDOI{\tempurl}


\bibitem[Hehn et~al\mbox{.}(2018)]%
        {HehnDT4RE}
\bibfield{author}{\bibinfo{person}{Jennifer Hehn}, \bibinfo{person}{Falk Uebernickel}, {and} \bibinfo{person}{Daniel Mendez~Fernandez}.} \bibinfo{year}{2018}\natexlab{}.
\newblock \showarticletitle{{DT4RE: Design Thinking for Requirements Engineering: A Tutorial on Human-Centered and Structured Requirements Elicitation}}. In \bibinfo{booktitle}{\emph{2018 IEEE 26th International Requirements Engineering Conference (RE)}}. \bibinfo{pages}{504--505}.
\newblock
\urldef\tempurl%
\url{https://doi.org/10.1109/RE.2018.00074}
\showDOI{\tempurl}


\bibitem[Heyn et~al\mbox{.}(2021)]%
        {heyn2021requirement}
\bibfield{author}{\bibinfo{person}{Hans-Martin Heyn}, \bibinfo{person}{Eric Knauss}, \bibinfo{person}{Amna~Pir Muhammad}, \bibinfo{person}{Olof Eriksson}, \bibinfo{person}{Jennifer Linder}, \bibinfo{person}{Padmini Subbiah}, \bibinfo{person}{Shameer~Kumar Pradhan}, {and} \bibinfo{person}{Sagar Tungal}.} \bibinfo{year}{2021}\natexlab{}.
\newblock \bibinfo{title}{Requirement Engineering Challenges for AI-intense Systems Development}.
\newblock
\newblock
\showeprint[arxiv]{2103.10270}~[cs.LG]


\bibitem[Hu et~al\mbox{.}(2020)]%
        {hu2020towards}
\bibfield{author}{\bibinfo{person}{Boyue~Caroline Hu}, \bibinfo{person}{Rick Salay}, \bibinfo{person}{Krzysztof Czarnecki}, \bibinfo{person}{Mona Rahimi}, \bibinfo{person}{Gehan Selim}, {and} \bibinfo{person}{Marsha Chechik}.} \bibinfo{year}{2020}\natexlab{}.
\newblock \showarticletitle{Towards requirements specification for machine-learned perception based on human performance}. In \bibinfo{booktitle}{\emph{2020 IEEE Seventh International Workshop on Artificial Intelligence for Requirements Engineering (AIRE)}}. IEEE, \bibinfo{pages}{48--51}.
\newblock


\bibitem[Ishikawa and Yoshioka(2019)]%
        {IshikawaYoshioka}
\bibfield{author}{\bibinfo{person}{Fuyuki Ishikawa} {and} \bibinfo{person}{Nobukazu Yoshioka}.} \bibinfo{year}{2019}\natexlab{}.
\newblock \showarticletitle{How Do Engineers Perceive Difficulties in Engineering of Machine-Learning Systems? - Questionnaire Survey}. In \bibinfo{booktitle}{\emph{2019 IEEE/ACM Joint 7th International Workshop on Conducting Empirical Studies in Industry (CESI) and 6th International Workshop on Software Engineering Research and Industrial Practice (SER\&IP)}}. \bibinfo{pages}{2--9}.
\newblock
\urldef\tempurl%
\url{https://doi.org/10.1109/CESSER-IP.2019.00009}
\showDOI{\tempurl}


\bibitem[ISO(2017)]%
        {iso12207}
\bibfield{author}{\bibinfo{person}{ISO}.} \bibinfo{year}{2017}\natexlab{}.
\newblock \bibinfo{title}{ISO/IEC/IEEE International Standard - Systems and Software Engineering -- Software Life Cycle Processes}.
\newblock , \bibinfo{numpages}{157}~pages.
\newblock
\urldef\tempurl%
\url{https://doi.org/10.1109/IEEESTD.2017.8100771}
\showDOI{\tempurl}


\bibitem[Jiang et~al\mbox{.}(2017)]%
        {jiang2017artificial}
\bibfield{author}{\bibinfo{person}{Fei Jiang}, \bibinfo{person}{Yong Jiang}, \bibinfo{person}{Hui Zhi}, \bibinfo{person}{Yi Dong}, \bibinfo{person}{Hao Li}, \bibinfo{person}{Sufeng Ma}, \bibinfo{person}{Yilong Wang}, \bibinfo{person}{Qiang Dong}, \bibinfo{person}{Haipeng Shen}, {and} \bibinfo{person}{Yongjun Wang}.} \bibinfo{year}{2017}\natexlab{}.
\newblock \showarticletitle{Artificial intelligence in healthcare: past, present and future}.
\newblock \bibinfo{journal}{\emph{Stroke and vascular neurology}} \bibinfo{volume}{2}, \bibinfo{number}{4} (\bibinfo{year}{2017}).
\newblock


\bibitem[Ku{\v{c}}ak et~al\mbox{.}(2018)]%
        {kuvcak2018machine}
\bibfield{author}{\bibinfo{person}{Danijel Ku{\v{c}}ak}, \bibinfo{person}{Vedran Juri{\v{c}}i{\'c}}, {and} \bibinfo{person}{Goran {\DJ}ambi{\'c}}.} \bibinfo{year}{2018}\natexlab{}.
\newblock \showarticletitle{MACHINE LEARNING IN EDUCATION-A SURVEY OF CURRENT RESEARCH TRENDS.}
\newblock \bibinfo{journal}{\emph{Annals of DAAAM \& Proceedings}}  \bibinfo{volume}{29} (\bibinfo{year}{2018}).
\newblock


\bibitem[Marban et~al\mbox{.}(2009)]%
        {Marban09}
\bibfield{author}{\bibinfo{person}{Oscar Marban}, \bibinfo{person}{Gonzalo Mariscal}, {and} \bibinfo{person}{Javier Segovia}.} \bibinfo{year}{2009}\natexlab{}.
\newblock \showarticletitle{A Data Mining \& Knowledge Discovery Process Model}.
\newblock In \bibinfo{booktitle}{\emph{Data Mining and Knowledge Discovery in Real Life Applications}}, \bibfield{editor}{\bibinfo{person}{Julio Ponce} {and} \bibinfo{person}{Adem Karahoca}} (Eds.). \bibinfo{publisher}{IntechOpen}, \bibinfo{address}{Rijeka}, Chapter~1.
\newblock
\urldef\tempurl%
\url{https://doi.org/10.5772/6438}
\showDOI{\tempurl}


\bibitem[Martins et~al\mbox{.}(2024a)]%
        {Martinscibse}
\bibfield{author}{\bibinfo{person}{Mariana Martins}, \bibinfo{person}{Taciana Kudo}, {and} \bibinfo{person}{Renato Bulcão-Neto}.} \bibinfo{year}{2024}\natexlab{a}.
\newblock \showarticletitle{A Qualitative Study on Requirements Engineering Practices in an Artificial Intelligence Unit of the Brazilian Industrial Research and Innovation Company}. In \bibinfo{booktitle}{\emph{Anais do XXVII Congresso Ibero-Americano em Engenharia de Software}} (Curitiba/PR). \bibinfo{publisher}{SBC}, \bibinfo{address}{Porto Alegre, RS, Brasil}, \bibinfo{pages}{46--60}.
\newblock
\showISSN{0000-0000}
\urldef\tempurl%
\url{https://sol.sbc.org.br/index.php/cibse/article/view/28438}
\showURL{%
\tempurl}


\bibitem[Martins et~al\mbox{.}(2024b)]%
        {MartinsWER}
\bibfield{author}{\bibinfo{person}{Mariana Martins}, \bibinfo{person}{Taciana Kudo}, {and} \bibinfo{person}{Renato Bulcão-Neto}.} \bibinfo{year}{2024}\natexlab{b}.
\newblock \showarticletitle{A Requirements Engineering Process for Machine Learning Innovation Projects}. In \bibinfo{booktitle}{\emph{Anais do XXVII Workshop de Engenharia de Requisitos}} (Buenos Aires/ARG). \bibinfo{publisher}{Even3, Brazil}, \bibinfo{address}{Buenos Aires, Argentina}.
\newblock
\urldef\tempurl%
\url{http://wer.inf.puc-rio.br/WERpapers/}
\showURL{%
\tempurl}


\bibitem[Martins et~al\mbox{.}(2025)]%
        {MartinsJSERD}
\bibfield{author}{\bibinfo{person}{Mariana~Crisostomo Martins}, \bibinfo{person}{Lívia Mancine~C. Campos}, \bibinfo{person}{João Lucas~R. Soares}, \bibinfo{person}{Taciana~Novo Kudo}, {and} \bibinfo{person}{Renato~F. Bulcão-Neto}.} \bibinfo{year}{2025}\natexlab{}.
\newblock \showarticletitle{Requirements Engineering for Machine Learning-Based AI Systems: A Tertiary Study}.
\newblock \bibinfo{journal}{\emph{Journal of Software Engineering Research and Development}} \bibinfo{volume}{13}, \bibinfo{number}{2} (\bibinfo{date}{Sep.} \bibinfo{year}{2025}), \bibinfo{pages}{13:129 – 13:142}.
\newblock
\urldef\tempurl%
\url{https://doi.org/10.5753/jserd.2025.4892}
\showDOI{\tempurl}


\bibitem[Nascimento et~al\mbox{.}(2020)]%
        {nascimento2020}
\bibfield{author}{\bibinfo{person}{Elizamary Nascimento}, \bibinfo{person}{Anh Nguyen-Duc}, \bibinfo{person}{Ingrid Sundbø}, {and} \bibinfo{person}{Tayana Conte}.} \bibinfo{year}{2020}\natexlab{}.
\newblock \bibinfo{title}{Software engineering for artificial intelligence and machine learning software: A systematic literature review}.
\newblock
\newblock
\showeprint[arxiv]{2011.03751}~[cs.SE]
\urldef\tempurl%
\url{https://arxiv.org/abs/2011.03751}
\showURL{%
\tempurl}


\bibitem[Pereira et~al\mbox{.}(2021)]%
        {conteBenefitsChallengesDT}
\bibfield{author}{\bibinfo{person}{Lauriane Pereira} {et~al\mbox{.}}} \bibinfo{year}{2021}\natexlab{}.
\newblock \showarticletitle{Towards an understanding of benefits and challenges in the use of design thinking in requirements engineering}. \bibinfo{pages}{1338--1345}.
\newblock
\urldef\tempurl%
\url{https://doi.org/10.1145/3412841.3442008}
\showDOI{\tempurl}


\bibitem[Perez(2017)]%
        {perez}
\bibfield{author}{\bibinfo{person}{Carlos~E. Perez}.} \bibinfo{year}{2017}\natexlab{}.
\newblock \bibinfo{booktitle}{\emph{The Deep Learning Ai Playbook: Strategy for Disruptive Artificial Intelligence}}.
\newblock \bibinfo{publisher}{Intuition Machine}.
\newblock


\bibitem[Pfleeger({[n.\,d.]})]%
        {pfleeger}
\bibfield{author}{\bibinfo{person}{S.L Pfleeger}.} \bibinfo{year}{[n.\,d.]}\natexlab{}.
\newblock \showarticletitle{Understanding and improving technology transfer in software engineering}.
\newblock \bibinfo{journal}{\emph{Journal of Systems and Software}} \bibinfo{volume}{47}, \bibinfo{number}{2} (\bibinfo{year}{[n.\,d.]}), \bibinfo{pages}{111--124}.
\newblock
\showISSN{0164-212}
\urldef\tempurl%
\url{https://doi.org/10.1016/S0164-1212(99)00031-X}
\showDOI{\tempurl}


\bibitem[Romao et~al\mbox{.}(2024)]%
        {romao2024sbes}
\bibfield{author}{\bibinfo{person}{Lucas Romao}, \bibinfo{person}{Marcos Kalinowski}, \bibinfo{person}{Clarissa Barbosa}, \bibinfo{person}{Allysson~Allex Araújo}, \bibinfo{person}{Simone D.~J. Barbosa}, {and} \bibinfo{person}{Helio Lopes}.} \bibinfo{year}{2024}\natexlab{}.
\newblock \bibinfo{title}{Agile Minds, Innovative Solutions, and Industry-Academia Collaboration: Lean R\&D Meets Problem-Based Learning in Software Engineering Education}.
\newblock
\newblock
\showeprint[arxiv]{2407.15982}~[cs.SE]
\urldef\tempurl%
\url{https://arxiv.org/abs/2407.15982}
\showURL{%
\tempurl}


\bibitem[Runeson and H\"{o}st(2008)]%
        {Runeson2008}
\bibfield{author}{\bibinfo{person}{Per Runeson} {and} \bibinfo{person}{Martin H\"{o}st}.} \bibinfo{year}{2008}\natexlab{}.
\newblock \showarticletitle{Guidelines for conducting and reporting case study research in software engineering}.
\newblock \bibinfo{journal}{\emph{Empirical Software Engineering}} \bibinfo{volume}{14}, \bibinfo{number}{2} (\bibinfo{date}{Dec.} \bibinfo{year}{2008}), \bibinfo{pages}{131–164}.
\newblock
\showISSN{1573-7616}
\urldef\tempurl%
\url{https://doi.org/10.1007/s10664-008-9102-8}
\showDOI{\tempurl}


\bibitem[Venkatesh and Bala(2008)]%
        {tam3}
\bibfield{author}{\bibinfo{person}{Viswanath Venkatesh} {and} \bibinfo{person}{Hillol Bala}.} \bibinfo{year}{2008}\natexlab{}.
\newblock \showarticletitle{Technology Acceptance Model 3 and a Research Agenda on Interventions}.
\newblock \bibinfo{journal}{\emph{Decision Sciences}} \bibinfo{volume}{39}, \bibinfo{number}{2} (\bibinfo{date}{May} \bibinfo{year}{2008}), \bibinfo{pages}{273–315}.
\newblock
\showISSN{1540-5915}
\urldef\tempurl%
\url{https://doi.org/10.1111/j.1540-5915.2008.00192.x}
\showDOI{\tempurl}


\bibitem[Villamizar et~al\mbox{.}(2021)]%
        {villamizar2021requirements}
\bibfield{author}{\bibinfo{person}{Hugo Villamizar}, \bibinfo{person}{Tatiana Escovedo}, {and} \bibinfo{person}{Marcos Kalinowski}.} \bibinfo{year}{2021}\natexlab{}.
\newblock \showarticletitle{Requirements engineering for machine learning: A systematic mapping study}. In \bibinfo{booktitle}{\emph{2021 47th Euromicro Conference on Software Engineering and Advanced Applications (SEAA)}}. IEEE, \bibinfo{pages}{29--36}.
\newblock


\bibitem[Villamizar et~al\mbox{.}(2022)]%
        {villamizar2022perspectivebased}
\bibfield{author}{\bibinfo{person}{Hugo Villamizar}, \bibinfo{person}{Marcos Kalinowski}, {and} \bibinfo{person}{Helio Lopes}.} \bibinfo{year}{2022}\natexlab{}.
\newblock \bibinfo{title}{Towards Perspective-Based Specification of Machine Learning-Enabled Systems}.
\newblock
\newblock
\showeprint[arxiv]{2206.09760}~[cs.SE]


\bibitem[Villamizar et~al\mbox{.}(2024)]%
        {villamizar2024identifying}
\bibfield{author}{\bibinfo{person}{Hugo Villamizar}, \bibinfo{person}{Marcos Kalinowski}, \bibinfo{person}{Hélio Lopes}, {and} \bibinfo{person}{Daniel Mendez}.} \bibinfo{year}{2024}\natexlab{}.
\newblock \showarticletitle{Identifying concerns when specifying machine learning-enabled systems: A perspective-based approach}.
\newblock \bibinfo{journal}{\emph{Journal of Systems and Software}}  \bibinfo{volume}{213} (\bibinfo{year}{2024}), \bibinfo{pages}{112053}.
\newblock
\showISSN{0164-1212}
\urldef\tempurl%
\url{https://doi.org/10.1016/j.jss.2024.112053}
\showDOI{\tempurl}


\bibitem[Villamizar et~al\mbox{.}(2018)]%
        {villamizarMapeamento}
\bibfield{author}{\bibinfo{person}{Hugo Villamizar}, \bibinfo{person}{Marcos Kalinowski}, \bibinfo{person}{Marx Viana}, {and} \bibinfo{person}{Daniel~Méndez Fernández}.} \bibinfo{year}{2018}\natexlab{}.
\newblock \showarticletitle{A Systematic Mapping Study on Security in Agile Requirements Engineering}. In \bibinfo{booktitle}{\emph{2018 44th Euromicro Conference on Software Engineering and Advanced Applications (SEAA)}}. \bibinfo{pages}{454--461}.
\newblock
\urldef\tempurl%
\url{https://doi.org/10.1109/SEAA.2018.00080}
\showDOI{\tempurl}


\bibitem[Vogelsang and Borg(2019)]%
        {vogelsang2019requirements}
\bibfield{author}{\bibinfo{person}{Andreas Vogelsang} {and} \bibinfo{person}{Markus Borg}.} \bibinfo{year}{2019}\natexlab{}.
\newblock \bibinfo{title}{Requirements Engineering for Machine Learning: Perspectives from Data Scientists}.
\newblock
\newblock
\showeprint[arxiv]{1908.04674}~[cs.LG]


\bibitem[Wan et~al\mbox{.}(2020)]%
        {Wan2020}
\bibfield{author}{\bibinfo{person}{Zhiyuan Wan}, \bibinfo{person}{Xin Xia}, \bibinfo{person}{David Lo}, {and} \bibinfo{person}{Gail~C. Murphy}.} \bibinfo{year}{2020}\natexlab{}.
\newblock \showarticletitle{How does Machine Learning Change Software Development Practices?}
\newblock \bibinfo{journal}{\emph{IEEE Transactions on Software Engineering}} (\bibinfo{year}{2020}), \bibinfo{pages}{1–1}.
\newblock
\showISSN{2326-3881}
\urldef\tempurl%
\url{https://doi.org/10.1109/tse.2019.2937083}
\showDOI{\tempurl}


\bibitem[Zaharia et~al\mbox{.}(2018)]%
        {zaharia2018accelerating}
\bibfield{author}{\bibinfo{person}{Matei Zaharia}, \bibinfo{person}{Andrew Chen}, \bibinfo{person}{Aaron Davidson}, \bibinfo{person}{Ali Ghodsi}, \bibinfo{person}{Sue~Ann Hong}, \bibinfo{person}{Andy Konwinski}, \bibinfo{person}{Siddharth Murching}, \bibinfo{person}{Tomas Nykodym}, \bibinfo{person}{Paul Ogilvie}, \bibinfo{person}{Mani Parkhe}, {et~al\mbox{.}}} \bibinfo{year}{2018}\natexlab{}.
\newblock \showarticletitle{Accelerating the machine learning lifecycle with MLflow.}
\newblock \bibinfo{journal}{\emph{IEEE Data Eng. Bull.}} \bibinfo{volume}{41}, \bibinfo{number}{4} (\bibinfo{year}{2018}), \bibinfo{pages}{39--45}.
\newblock


\end{thebibliography}

\end{document}